\documentclass[aps,prc,tightenlines,showpacs,superscriptaddress,twocolumn]
{revtex4}

\usepackage{graphicx,color}

\begin{document}

\title{Isospin Diffusion in Heavy-Ion Collisions and 
the Neutron Skin Thickness of Lead}

\author{Andrew W. Steiner} 
\affiliation{Theoretical Division, Los Alamos National Laboratory,
Los Alamos, NM 87545, USA}

\author{Bao-An Li} 
\affiliation{Department of Chemistry and Physics, P.O. Box 419,
Arkansas State University, State University, Arkansas, 72467-0419, USA}

\begin{abstract} 
The correlation between the thickness of the neutron skin in
$^{208}$Pb, and the degree of isospin diffusion in heavy-ion
collisions is examined. The same equation of state is used to compute
the degree of isospin diffusion in an isospin-depedent transport model
and the neutron skin thickness in the Hartree-Fock approximation. We
find that skin thicknesses less than 0.15 fm are excluded by the
isospin diffusion data. 
\end{abstract}

\pacs{25.70.-z, 21.30.Fe, 21.10.Gv, 24.10.Lx}

\maketitle

\section{Introduction}

The nuclear symmetry energy and its dependence on density holds a
unique place in nuclear physics and astrophysics. It is an essential
ingredient in understanding many aspects of nuclear physics, from
heavy-ion collisions to nuclear structure, and astrophysics, from
supernovae and neutron stars to r-process nucleosynthesis (for a
recent review see Ref.~\cite{Steiner05}). At the same time, the
nuclear symmetry energy is relatively uncertain; it is known to within
only about 5 MeV at saturation density (compared to the binding energy
which is known to within 1 MeV). The density dependence of the nuclear
symmetry energy at both sub-saturation and super-saturation densities
is very poorly known.

The symmetry energy is also an important quantity in determining the
compressibility of matter.  The incompressibility $K_{0}$ of symmetric
nuclear matter at its saturation density $\rho _{0}=0.16$ fm$^{-3}$
has been determined to be $231\pm 5$ MeV from nuclear giant monopole
resonances~\cite{youngblood99} and the equation of state (EOS) at
densities of $2\rho_0<\rho<5\rho_0$ has been constrained by
measurements of collective
flows~\cite{Gutbrod89,Danielewicz85,Gustafsson84,Welke88,Siemens79} in
relativistic heavy-ion collisions~\cite{Danielewicz02}. However,
further progress in determining more precisely both the parameter
$K_{0}$ and the EOS of symmetric nuclear matter is severely hindered
by the uncertainties of the symmetry
energy~\cite{Danielewicz02,Colo04}.

The possibility of determining the EOS of nuclear matter from
heavy-ion collisions has been discussed for almost 30 years, and
indeed impressive progress has been achieved. A number of heavy-ion
collision probes of the symmetry energy have been proposed including
isospin fractionation~\cite{Muller95,Li97,Baran98,Xu00,Tan01},
isoscaling~\cite{Tsang01,Ono03}, neutron-proton differential
collective flow~\cite{Li00}, pion production~\cite{Li02} and
neutron-proton correlation function~\cite{Chen03}.  Determination of
the equation of state from heavy-ion data is typically performed
through comparisons with transport model
simulations~\cite{Danielewicz02,Li98,Li01b,Baran05,Li05}. A recent
analysis of the symmetry energy comes from isospin diffusion data for
reactions involving $^{112}$Sn and $^{124}$Sn from the
NSCL/MSU~\cite{Tsang04,Chen05}. Using an isospin- and
momentum-dependent transport model, IBUU04~\cite{Li04,Li04b}, the
NSCL/MSU data on isospin diffusion were found to be consistent with a
nuclear symmetry energy of $E_{sym}(\rho )\approx 31.6(\rho
/\rho_{0})^{1.05}$ at sub-saturation densities. 

The pressure of neutron-rich matter at densities near two-thirds
saturation density (which is closely related to the symmetry energy,
see Ref.~\cite{Steiner05}) is tightly correlated to the neutron skin
thickness, $\delta R$, in $^{208}$Pb~\cite{Brown00,Typel01}. The
present theoretical uncertainty in the symmetry energy results in a
large uncertainty in the theoretical calculations of the skin
thickness. Experimental measurements of the skin thickness are also
very uncertain, mostly resulting from the difficulty in disentangling
the strong interactions between the probe and the neutrons in the
nucleus. Values from 0.07 fm to 0.24 fm have been suggested by the
data~\cite{Starodubsky94,Clark03}. A determination of the symmetry
energy at sub-saturation densities would offer a prediction of skin
thickness of lead or vice versa. 

The structure of nuclei and neutron stars and the properties of
heavy-ion collisions are all determined by the same underlying EOS. In
particular, both the size of the neutron skin and the degree of
isospin diffusion in heavy-ion collisions at intermediate energies are
sensitive to the symmetry energy at sub-saturation densities. In this
article, we study the correlation between the thickness of neutron
skin in $^{208}$Pb and the strength of isospin diffusion using the
same set of equations of state for asymmetric nuclear matter. This
differs from previous work in that, for the first time, we make a
direct connection between observables from nuclear structure and
heavy-ion collisions. The transport model IBUU04 was used to predict
the isospin diffusion for several equations of state which differ only
in the density dependence of the symmetry energy. These equations of
state are also used to calculate $\delta R$ for $^{208}$Pb in the
Hartree-Fock (HF) approximation. The correlation between the $\delta
R$ and the isospin diffusion provides a more stringent constraint than
their individual values on the EOS of neutron-rich matter. Comparisons
with the available data are also discussed.

\section{The Equation of State and The Transport Model Calculations of
Isospin Diffusion in Heavy-Ion Collisions}

Isospin diffusion measures quantitatively the net exchange of 
isospin contents between the projectile and target nuclei.  
Using symmetric reactions $A+A$ and $B+B$ as references the 
degree of isospin diffusion in the asymmetric reaction of $A+B$ 
can be measured by~\cite{Rami00}
\begin{equation}
R_i=\frac{2X^{A+B}-X^{A+A}-X^{B+B}}{X^{A+A}-X^{B+B}}\label{Ri}
\end{equation}
where $X$ is any isospin-sensitive tracer.  In the recent NSCL/MSU
experiments with $^{124}$Sn on $^{112}$Sn at a beam energy of 50
MeV/nucleon and an impact parameter about 6 fm, the isospin asymmetry
of the projectile-like residue was used as the isospin
tracer~\cite{Tsang04}. Consistent with the experimental selection, in
model analyses the average isospin asymmetry $\left\langle \delta
\right\rangle $ of the $^{124}$Sn-like residue was calculated from
nucleons with local densities higher than $\rho_{0}/20$ and velocities
larger than $1/2$ the beam velocity in the c.m. frame.  Reactions at
intermediate energies are always complicated by preequilibrium
particle emission and the production of possibly neutron-rich
fragments at mid-rapidity, however, the quantity $R_i$ has the
advantage of minimizing significantly these effects~\cite{Tsang04}.

The NSCL/MSU data were recently analyzed by using the IBUU04
version of an isospin and momentum dependent transport
model~\cite{Li04,Li04b}. Here, we recall the ingredients of the model
that are most relevant for the present study. In this model one can
select to use either the experimental nucleon-nucleon cross sections
in free space or reduced in-medium nucleon-nucleon cross
sections. Since the later involves several model-dependent
assumptions, we have used the free-space cross sections (as in
Ref.~\cite{Chen05}).  The EOS used in this model is based on the Gogny
effective interactions. The potential $U(\rho, \delta,
\mathbf{p}, \tau)$ for a nucleon with isospin $\tau $ ($1/2$ for
neutrons and $-1/2$ for protons) and momentum $\mathbf{p}$ in
asymmetric nuclear matter at total density $\rho $ is given
by~\cite{Prakash88b,Prakash97,Bombaci01,Das03}
\begin{eqnarray}
U_{\text{MDI}}(\rho ,\delta ,\mathbf{p},\tau ) &=&A_{u}\frac{\rho _{\tau
^{\prime }}}{\rho _{0}}+A_{l}\frac{\rho _{\tau }}{\rho _{0}}+B\left( \frac{%
\rho }{\rho _{0}}\right) ^{\sigma }(1-x\delta ^{2})  \nonumber \\
&-&8\tau x\frac{B}{\sigma +1}\frac{\rho ^{\sigma -1}}{\rho _{0}^{\sigma }}%
\delta \rho _{\tau ^{\prime }}  \nonumber \\
&+&\frac{2C_{\tau ,\tau }}{\rho _{0}}\int d^{3}\mathbf{p}^{\prime }\frac{%
f_{\tau }(\mathbf{r},\mathbf{p}^{\prime })}{1+(\mathbf{p}-\mathbf{p}^{\prime
})^{2}/\Lambda ^{2}}  \nonumber \\
&+&\frac{2C_{\tau ,\tau ^{\prime }}}{\rho _{0}}\int d^{3}\mathbf{p}^{\prime }%
\frac{f_{\tau ^{\prime }}(\mathbf{r},\mathbf{p}^{\prime })}{1+(\mathbf{p}-%
\mathbf{p}^{\prime })^{2}/\Lambda ^{2}},  \label{mdi}
\end{eqnarray}%
where $\rho _{\tau }$ and $\rho _{\tau }^{\prime }$ denote proton or
neutron density with $\tau \neq \tau ^{\prime }$; and $\delta \equiv
(\rho _{n}-\rho _{p})/\rho $ is the isospin asymmetry. The $f_{\tau
}(\mathbf{r},\mathbf{p})$ denotes the phase-space distribution
function at coordinate $\mathbf{r}$ and momentum $\mathbf{p}$. The
corresponding momentum-dependent interaction (MDI) leads to an
incompressibility of $K_{0}=211$ MeV for the symmetric nuclear matter
at saturation density. On the right hand side of Eq.(\ref{mdi}), the
first four terms with $\sigma =3/4$ and $B=106.35$ MeV describe the
momentum-independent interaction. The terms with parameters $C_{\tau
,\tau }=-11.7$ MeV and $C_{\tau ,\tau ^{\prime }}=-103.4$ MeV describe
the momentum-dependent interaction of a nucleon of isospin $\tau $ and
momentum $\mathbf{p}$ with like and unlike nucleons in the background
fields, respectively. We stress that the momentum dependence in both
the isoscalar and isovector potentials is taken into account in a way
consistent with that known empirically from nucleon optical
potentials~\cite{Li04a}. With the parameter $\Lambda=1.0p_{F}^{0}$,
where $p_{F}^{0}$ denotes nucleon Fermi momentum at $\rho _{0}$, the
isoscalar potential $(U_{n}(\rho,p)+U_{p}(\rho,p))/2$ coincides with
predictions from the variational many-body theory using inputs
constrained by nucleon-nucleon scattering data~\cite{Wiringa88,Li04a},
and the isovector potential $(U_{n}(\rho ,p)-U_{p}(\rho ,p))/2$ also
agrees with the momentum dependence of the Lane potential extracted
from nucleon-nucleus scattering experiments up to about 100
MeV~\cite{Hodgson94} and (p,n) charge exchange reactions up to about 45
MeV~\cite{Hoffmann72,Li04a}.

\begin{figure}
\includegraphics[angle=270,scale=0.4]{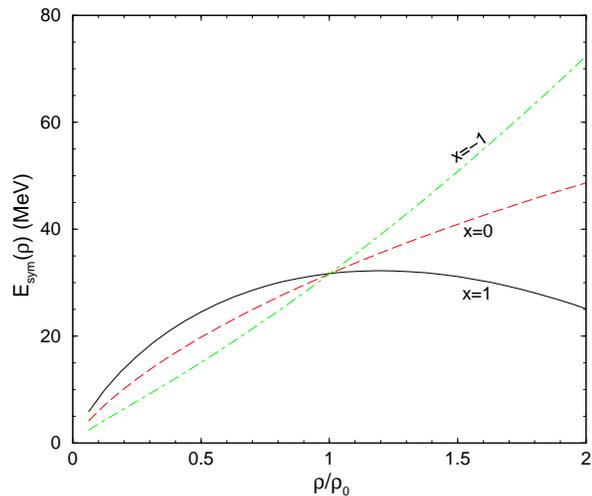}
\caption{The symmetry energies as a function of baryon number density,
$\rho$, in units of the saturation density, $\rho_0$, for three of the
equations of state considered in this work.}
\label{fig:esym-x}
\end{figure}

The parameter $x$ in Eq.(\ref{mdi}) is introduced to allow variations
in the density dependence of the nuclear symmetry energy
$E_{\text{sym}}(\rho )$, which is defined via the parabolic
approximation to the nucleon specific energy in isospin asymmetric
nuclear matter~\cite{Li98,Li01b}, i.e.,
\begin{equation}
E(\rho ,\delta )=E(\rho ,\delta =0)+E_{\text{sym}}(\rho )\delta ^{2}+%
\mathcal{O}(\delta ^{4}).  \label{eos}
\end{equation}%
The symmetry energies for three of the models considered here are
displayed in Fig.~\ref{fig:esym-x} as a function of density.  With
$x=1$, $E_{\text{sym}}(\rho)$ matches what is predicted by a
Hartree-Fock calculation using the Gogny effective
interaction~\cite{Das03}. The parameters $A_{l}(x)$ and $A_{u}(x)$ are
$A_{l}(x)=-120.57+2Bx/(\sigma +1)$ and $A_{u}(x)=-95.98-2Bx/(\sigma
+1)$, respectively. It is important to note that large positive
(negative) values of $x$ lead to a ``higher'' (``lower'') symmetry
energy at sub-saturation densities, but a lower (higher) symmetry
energy at super-saturation densities.

\begin{figure}
\includegraphics[scale=0.8]{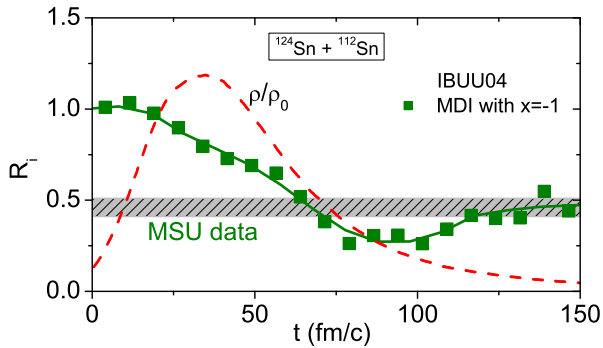}
\caption{Time evolutions of $R_i$ and average central density
for MDI and SBKD interactions with $x=-1$.}
\label{fig:rtime}
\end{figure}

An example of the results from the transport calculations, taken from
Ref.~\cite{Chen05} is given in Fig.~\ref{fig:rtime} for the MDI EOS
with $x=-1$. The time evolution of $R_i$ is given for this equation of
state. The evolution of the average central density $\rho$ in units of
the saturation density $\rho_0$ is also given. It is clear that the
late-time evolution of $R_i$ for the MDI equation of state matches the
data obtained from MSU. The central density at these times is quite
small, indicating clearly that $R_i$ is sensitive to the symmetry
energy at sub-saturation densities.

\section{The EOS Fitting Procedure and Properties of $^{208}$Pb}

\begin{figure}
\includegraphics[scale=0.4]{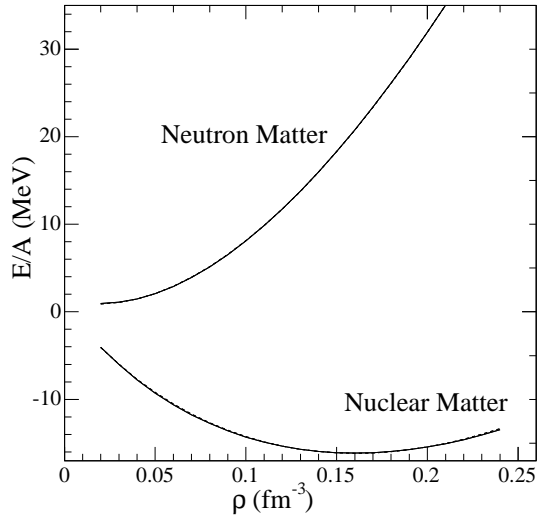}
\caption{The results of fitting the MDI EOS for $x=-1$ (solid lines) to
a Skyrme EOS (dashed lines). Plotted is the energy per baryon as a
function of the baryon density, $\rho$. The dashed lines are barely
visible because of the high quality of the fit.}
\label{fig:fit}
\end{figure}

Because the MDI EOS in the form above does not easily avail itself to
the calculation of the structure of lead nuclei we choose to fit the
EOS to a Skyrme model and use a Skyrme-Hartree-Fock code to calculate
corresponding lead nucleus. This procedure has been used
successfully to predict the neutron skin based on the Akmal,
et. al. EOS~\cite{Akmal98} in Ref.~\cite{Steiner05}. The four MDI
equations of state with different values of $x$ were fit to Skyrme
models constrained so values for the binding energy ($E_{b}$) and
charge radius ($R_{ch}$) for lead which were within 2\% of the
experimental values. For the equation of state with $x=-1$, the
results of the fit are displayed in Fig.~\ref{fig:fit}. The similar
nature of the Skyrme and MDI parameterizations allows for a very close
fit. For all of the four fits employed, the slope, $L\equiv
3\rho_0\left(dE_{\text{sym}}/d\rho\right)_{\rho=\rho_0}$, and the
curvature $K_{\text{sym}}\equiv 9\rho_0^{2}
\left(d^{2}E_{\text{sym}}/d\rho ^{2}\right)_{\rho=\rho_0}$ of the
symmetry energy and the final properties of lead nuclei in the
Hartree-Fock approximation are given in Table~\ref{tab:fit}. Also
given in the table are the experimental values. 

The effective mass for the original MDI EOS and the new Skyrme model
are not as close, for example, the effectve mass at saturation density
for the MDI EOS with $x=-1$ is 0.67, while for the Skyrme fit, it is
0.77. However, the neutron skin thickness is more sensitive to the
bulk energetics than the effective mass. As a demonstration, a new
Skyrme fit to the MDI EOS with $x=-1$ with a lower effective mass of
0.65 only lowers the skin thickness by 0.01 fm.

\begin{table}
\begin{tabular}{cccccc}
x & $E_{b}$ (MeV) & $R_{ch}$ (fm) & $\delta R$ (fm) & L (MeV) & 
$K_{\mathrm{sym}}$ (MeV) \\
\hline
-2 & 7.8598 & 5.517 & 0.38 & 153.0 & 276.3\\
-1 & 7.8702 & 5.520 & 0.28 & 107.4 & 94.1 \\
0 & 7.8584 & 5.547 & 0.22  & 62.1  & -88.6\\
1 & 7.8719 & 5.522 & 0.16  & 16.4  & -270.4\\
\hline
Exp. & 7.8675 & 5.503 & 0.07-0.24 & & \\
\end{tabular}
\caption{The properties of lead nuclei for the Skyrme equations
of state and their experimental values. $L$ and $K_{sym}$ are the 
slope and curvature of the corresponding symmetry energy at $\rho_0$.}
\label{tab:fit}
\end{table}

\section{Isospin Diffusion versus the Neutron-Skin of $^{208}$Pb }
Now we turn to the correlation between the degree of isospin diffusion
and the size of neutron-skin in $^{208}$Pb. Do we expect a correlation
between these two seemingly different observables?  As it has been
discussed in detail in Refs.~\cite{Horowitz01,Horowitz02}, the size of
neutron skins in heavy nuclei is determined by the difference in
pressures on neutrons and protons $\delta P$. The latter is
propotional to $L$, which measures the stiffness of the symmetry enrgy
at saturation density. As shown in Table~\ref{tab:fit}, the
symmetry energy becomes more stiff (for densities larger than the
saturation density) as $x$ decreases from 1 to $-2$, and
the size of the neutron skin in $^{208}$Pb increases. Alternatively,
one can understand this effect by noting that the the symmetry energy
is lower (e.g., with $x=-2$) at subsaturation densities the cost of
creating a difference between the neutron and proton densities is
small and the skin thickness is large. As the symmetry energy
increases with larger values of $x$, the neutron skin thickness
decreases accordingly.

\begin{figure}
\includegraphics[scale=0.4]{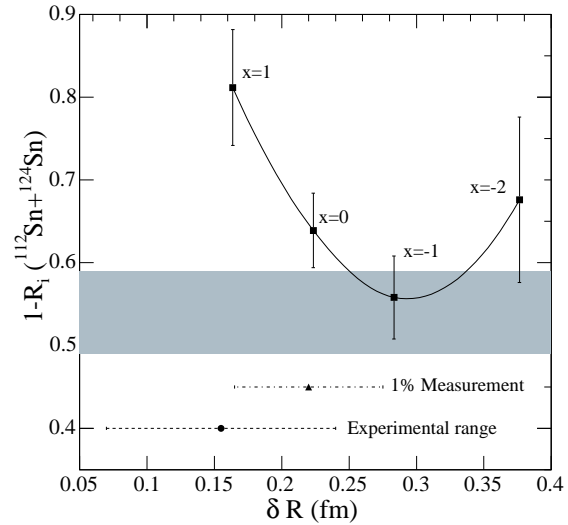}
\caption{The skin thickness in lead, $\delta R$, versus the isospin
diffusion parameter $1-R_i$ for the four equations of state in this
work. A sample measurement with an uncertainty corresponding to a 1\%
measurement of the neutron radius is included, as well as the present
acceptable experimental range.}
\label{fig:rdr}
\end{figure}

It is also well known that the degree of isospin diffusion in
heavy-ion collisions depends sensitively on the stiffness of the
symmetry energy~\cite{Shi03,Tsang04,Li04c,Chen05,Baran05}. A
correlation between the degree of isospin diffusion in heavy-ion
collisions and the size of neutron skins in heavy nuclei using the
same EOS is thus expected. Of course, it is understood that this
correlation is not universal as both the isospin diffusion and size of
neutron skins are system dependent. However, an examination of this
correlation for any system and {\it simultaneous} comparisons with the
corresponding data may provide a more stringent constraint on the
underlying EOS. Shown in Fig.~\ref{fig:rdr} is our analysis of this
correlation for the isospin diffusion in $^{124}$Sn+$^{112}$Sn and the
neutron skin in $^{208}$Pb. The available data on both quantities are
also included in the graph. There are some remaining systematic
uncertainties that are difficult to estimate which are not addressed
here (see Ref.~\cite{Li04} for more details).  

From Eq.~\ref{Ri}, we can see that $R_i=1$ implies the projectile-like
residue has the same isospin asymmetry as the projectile $^{124}$Sn
without any net exchange of isospin asymmetry with the target
$^{112}$Sn.  This is also often referred as the complete isospin
transparency. A value of zero for $R_i$ implies that a complete mixing
between the projectile and target indicating the establishment of an
isospin equilibrium during the reaction.

The NSCL/MSU data indicates that the mixing is about 50\%. Unlike the
relation between $x$ and the neutron skin thickness, the relation
between $R_i$ and the parameter $x$ is non-monotonic.  This is because
the $R_i$ is directly related to the symmetry potential which depends
on both the density and momentum as discussed in detail in
Ref.~\cite{Chen05}. Transport model calculations with $x=-1$ are
closest to the experimental data, although $x=0$ and $x=-2$ are not
necessarily excluded within the current uncertainties of both the experimental
data and model calculations. 

Upon comparing with previous estimates for $\delta R$, the connection
with the moderate amount of isospin diffusion displayed in heavy-ion
reactions suggests that $\delta R$ is on the upper range of the
present experimental uncertainty. Only skin thicknesses larger than
about 0.2 fm are consistent with the transport model
calculations. Skin thicknesses lower than 0.15 fm appear to be
excluded. However, the isospin dependence of the in-medium
nucleon-nucleon cross sections may affect the isospin diffusion.  It
was shown analytically that the degree of isospin diffusion depends on
both the symmetry energy and the neutron-proton scattering cross
section~\cite{Shi03}.  Because of the momentum-dependence of the
single nucleon potential of eq.\ \ref{mdi}, not only the values of the
n-n and p-p cross sections are reduced and different from each other
compared to their free-space values, the ratio
$2\sigma_{np}/(\sigma_{pp}+\sigma_{nn})$ is also modified from their
free-space value. Whie the reduction of in-medium n-n and p-p
scattering cross sections is expected to have little effect on the
value of $R_i$, the reduced in-medium n-p scattering cross sections
leads to smaller isospin diffusion.  To reproduce the NSCL/MSU data a
larger symmetry energy at sub-saturation densities is then needed. The
use of in-medium cross sections will, in effect, decrease the
preferred neutron skin thicknesses for the isospin diffusion data. An
investigation of this effect is currently underway and the results
will be reported elsewhere.

Although isospin diffusion in heavy-ion collisions primarily
constrains the symmetry energy at sub-saturation densities, it is
useful to check to ensure that the information from heavy-ion
collisions is not inconsistent with neutron star observations. Neutron
star radii are sensitive probes of the symmetry energy at higher
density~\cite{Lattimer01}. We solve the Tolman-Oppenheimer-Volkov
equations for the MDI EOS with $x=0,-1$, and $-2$. The maximum masses
(in order of decreasing $x$) are 1.92, 2.08, and 2.13 $M_{\odot}$ and
the radii of the maximum mass stars are 10.1, 11.2, and 11.8 km.  In
addition to being consistent with the compactness constraint given by
the measurement of the redshift of EXO0748-676 in
Ref.~\cite{Cottam02}, this equation of state is also consistent with
the recent neutron star radius measurement in Ref.~\cite{Gendre03}.

Recently, Ref.~\cite{ToddRutel05} has also explored the impact of the
symmetry energy on the neutron skin thickness in lead. Using an
relativistic mean-field model they construct an equation of state
which faithfully describes the giant monopole and dipole resonances in
nuclei. Their model suggests a neutron skin thickness of 0.21 fm in
Pb$^{208}$ which is consistent with our constraints from isospin
diffusion in heavy-ion collisions. 

\section{Summary}

In summary, we examined the correlation of the degree of isospin
diffusion in heavy-ion collisions and the size of neutron skin in
$^{208}$Pb using the same EOS within the IBUU04 transport model and
the HF approach. We found that neutron skin thicknesses less than 0.15
are disfavored by the isospin diffusion data. Reduction of the
experimental and theoretical uncertainties will likely lead to an
important constraint on the symmetry energy at sub-saturation density.

We would like to thank Lie-Wen Chen and Madappa Prakash for helpful
discussions. This work was supported in part by the DOE under grant
no. DOE/W-7405-ENG-36 (for AWS) and by the NSF under grant
no. PHY-0243571, PHY-0354572 and the NASA-Arkansas Space Grant
Consortium Award ASU15154 (for B.-A. Li)

\bibliography{paper8}
\bibliographystyle{apsrev}
\end{document}